\documentclass[10pt,aps,pra,a4paper,australian,twocolumn]{revtex4-2}
\usepackage{lmodern}
\usepackage[T1]{fontenc}
\usepackage{babel}
\usepackage{float}
\usepackage{booktabs}
\usepackage{amsmath}
\usepackage{amssymb}
\usepackage[pdftex]{graphicx}

\usepackage{hyperref}

\hypersetup{
    unicode=false,          
    pdftoolbar=true,        
    pdfmenubar=true,        
    pdffitwindow=false,     
    pdfstartview={FitH},    
    pdfauthor={Sean Hodgman},     
    colorlinks=true,       
    linkcolor=blue,          
    citecolor=red,        
}

\makeatletter

\pdfpageheight\paperheight
\pdfpagewidth\paperwidth

\newcommand{\noun}[1]{\textsc{#1}}
\providecommand{\tabularnewline}{\\}

\usepackage{mathtools}
\usepackage{threeparttable}

\renewcommand\[{\begin{equation}}
\renewcommand\]{\end{equation}}

\DeclareUnicodeCharacter{03A3}{$\Sigma$}

\makeatother

\begin{document}
\title{Close-coupled model of Feshbach resonances in ultracold \texorpdfstring{$^{3}$He$^*$}{3He*} and \texorpdfstring{$^{4}$He$^*$}{4He*} atomic collisions}
\author{T. M. F. Hirsch}
\thanks{Present address: School of Mathematics and Physics, The University
of Queensland, St Lucia 4072, Australia}
\email{timothy.hirsch@uq.edu.au}

\affiliation{Research School of Physics, Australian National University, Canberra
0200, Australia}
\author{D. G. Cocks}
\affiliation{Research School of Physics, Australian National University, Canberra
0200, Australia}
\author{S. S. Hodgman}
\affiliation{Research School of Physics, Australian National University, Canberra
0200, Australia}
\begin{abstract}
Helium atoms in the metastable $2\prescript{3}{}{S_{1}}$ state (He$^*$)
have unique advantages for ultracold atomic experiments. However,
there is no known accessible Feshbach resonance in He$^*$, which could
be used to manipulate the scattering length and hence unlock several
new experimental possiblities. Previous experimental and theoretical
studies for He$^*$ have produced contradictory results. We aimed to
resolve this discrepancy with a theoretical search for Feshbach resonances,
using a new close-coupled model of He$^*$ collisions in the presence
of an external magnetic field. Several resonances were detected and
the existing literature discrepancy was resolved. Although none of
the resonances identified are readily experimentally useable, an interesting
non-Feshbach scattering length variation with magnetic field was observed
in heteronuclear collisions, at field strengths that are experimentally
accessible.
\end{abstract}
\maketitle

\section{Introduction}

Ultracold atomic gases are a unique experimental platform for the
study of atomic and molecular physics and quantum mechanics. The simplicity,
isolation and fewer degrees of freedom offered by such systems allow
exciting experiments to be performed such as Bose-Einstein condensation
\citep{andersonObservationBoseEinsteinCondensation1995} and the trapping
of atoms in periodic optical lattices \citep{blochManybodyPhysicsUltracold2008}.
One feature of ultracold gases is that the collisions between atoms
are accurately described by a single parameter---the s-wave scattering
length. Feshbach resonances allow the tuning of the scattering length,
typically via the application of an external magnetic field. Suitable
resonances have facilitated experiments such as exploring quantum
phase transitions \citep{romansQuantumPhaseTransition2004}, superfluidity
regimes \citep{blochQuantumSimulationsUltracold2012}, realising negative
temperature states \citep{braunNegativeAbsoluteTemperature2013},
or `Bose-nova' condensate collapse \citep{donleyDynamicsCollapsingExploding2001}.

Metastable helium (He$^*$) atoms possess unique advantages in ultracold
experiments as the large internal energies of these atoms allow charged
plate detectors to achieve individual atom detection with relatively
high spatial and temporal resolution \citep{vassenColdTrappedMetastable2012}.
He$^*$ atoms are trapped in the $2\prescript{3}{}{}S_{1}$ excited
state of helium, which has a lifetime of $2.2$ hours \citep{hodgmanMetastableHeliumNew2009},
longer than the timescale of typical ultracold experiments, and an
internal energy of $19.8\,\mathrm{eV}$ \citep{mortonEnergyLevelsStable2006},
the largest of any metastable atom. These properties have facilitated
unique experiments such as studies of Hanbury Brown-Twiss style matter
wave interference \citep{jeltesComparisonHanburyBrown2007}, production
and detection of Bell correlated atomic pairs \citep{shinBellCorrelationsSpatially2019,dussarratTwoParticleFourModeInterferometer2017}
and ghost imaging \citep{hodgmanHigherOrderQuantumGhost2019}. Several
extensions to these, as well as other proposed experiments for He$^*$,
would benefit from the use of a Feshbach resonance to tune the scattering
length. These include controlling the rate of entanglement generating
collisions \citep{perrinObservationAtomPairs2007,shinBellCorrelationsSpatially2019},
adjusting the two-body interaction parameter in many-body lattice
experiments \citep{caylaHanburyBrownTwiss2020} and increasing the
quantum depleted fraction of a BEC \citep{changMomentumResolvedObservationThermal2016,rossSurvivalQuantumDepletion2021}.

Unfortunately, He$^*$ experiments have been unable to use a Feshbach
resonance, because it is unknown if an experimentally accessible Feshbach
resonance exists in this state. There has been a single theoretical
study of this problem \citep{goosenFeshbachResonances3He2010} which
used a perturbative treatment called the Asymptotic Bound State model.
That study predicted multiple resonances over the different isotopic
mixtures, albeit many were too narrow and/or required too high magnetic
field strengths to be experimentally useful. However, an experimental
search for the most promising of the predicted $\prescript{4}{}{}$He$^*$
resonances failed to observe it \citep{borbelyMagneticfielddependentTrapLoss2012},
suggesting further investigation is required.

In this manuscript we develop a non-perturbative, ab-initio close
coupled model of metastable helium atoms in the presence of a magnetic
field. Our results resolve the discrepancy between the prior study
and experiment: the resonance that was not seen in the experiment
is broadened by ionisation effects that were discounted in the perturbative
treatment. We identify several other resonances, as well as interesting
non-Feshbach resonance magnetic field dependence.

\section{Methods}

This approach mostly follows that of \citep{cocksUltracoldHomonuclearHeteronuclear2019},
extending the method to include the Zeeman interaction. Briefly, we
write the scattering wavefunction of the colliding atoms in a basis
of states called channels, and use that decomposition to solve the
scattering Schr\"odinger equation. Fitting to the large-distance
solution yields a scattering matrix containing information such as
cross sections. The model was written in the \noun{Julia} programming
language \citep{bezansonJuliaFreshApproach2017} using the \noun{DifferentialEquations}
package \citep{rackauckasDifferentialEquationsJlPerformant2017}.

\subsection{Hamiltonian}

The scattering of two metastable helium atoms in an external magnetic
field is governed by the Hamiltonian:
\[
\hat{H}=\hat{T}+\hat{H}_{\mathrm{el}}+\hat{H}_{\mathrm{rot}}+\hat{H}_{\mathrm{sd}}+\hat{H}_{\mathrm{zee}}+\hat{H}_{\mathrm{hfs}},
\]
where
\[
\hat{T}=\frac{-\hbar^{2}}{2\mu}\frac{1}{R^{2}}\frac{\partial}{\partial R}\left(R^{2}\frac{\partial}{\partial R}\right)
\]
is the kinetic term, and 
\[
\hat{H}_{\mathrm{rot}}=\frac{\hat{l}^{2}}{2\mu R^{2}}
\]
is the centripetal operator effective in spherical coordinates.

\[
\hat{H}_{\mathrm{el}}=\hat{H}_{1}+\hat{H}_{2}+\hat{H}_{12}
\]
is the electronic Hamiltonian, $\hat{H}_{i}$ representing the kinetic
term for the electrons in atom $i$ and $\hat{H}_{12}$ representing
the Coulombic interaction between the two atoms.

\[
\hat{H}_{sd}=-\frac{\xi}{\hbar^{2}R^{3}}[3(\hat{\mathbf{S}}_{1}\cdot\mathbf{\hat{R}})(\mathbf{\hat{S}}_{2}\cdot\mathbf{\hat{R}})-\mathbf{\hat{S}}_{1}\cdot\mathbf{\hat{S}}_{2}]
\]
is the spin-dipole interaction between the magnetic dipole moments
of the two atoms. In the above, $\mathbf{\hat{S}}_{i}$ is the spin
operator for atom $i$, $\hat{\mathbf{R}}$ is the unit vector along
the internuclear axis, and
\[
\xi=\alpha^{2}\left(\frac{\mu_{e}}{\mu_{B}}\right)^{2}\,E_{\mathrm{h}}a_{0}^{3}.
\]
Here $\alpha$ is the fine structure constant, $\mu_{e}$ and $\mu_{B}$
are respectively the electron magnetic moment and Bohr magneton, $1\,E_{\mathrm{h}}\approx27.2\,\mathrm{eV}$
is a Hartree, and $1\,a_{0}\approx0.529\,\mathring{A}$ is a Bohr
radius.

\[
\hat{H}_{\mathrm{zee}}=-\frac{g_{s}\mu_{B}}{\hbar}\mathbf{B\cdot}(\mathbf{S}_{1}+\mathbf{S}_{2})
\]
is the Zeeman interaction of the atomic spins in the magnetic field,
where $g_{s}\approx2$ is the electronic spin g-factor.

Finally, $\hat{H}_{\mathrm{hfs}}$ is the hyperfine structure term
which is only for $\prescript{3}{}{}$He$^*$ atoms. The form of $\hat{H}_{\mathrm{hfs}}$
is given below in Section \ref{subsec:Matrix-elements}.

\subsection{Coupled channels approach}

We use the the coupled channel formalism, which expands the scattering
wavefunction $|\Psi\rangle$ into a basis of molecular states $|a\rangle$
using radial functions $G_{a}(R)$:
\[
|\Psi(R)\rangle=\sum\frac{1}{R}G_{a}(R)|a\rangle.
\]
The molecular states, called channels, are eigenstates of the system
Hamiltonian at asymptotically large separation distances. We label
them open or closed if the eigenvalues of 
$\hat{H}_{\mathrm{asymp}}=\hat{H}_{1}+\hat{H}_{2} + \hat{H}_{\mathrm{zee}}+\hat{H}_{\mathrm{hfs}}$ are lesser or greater, respectively, than the total energy E
(i.e. closed channels are energetically forbidden
at large separation distances). We also invoke the Born-Oppenheimer
approximation, assuming that the channels have only a parametric dependence
on the internuclear separation $R$. This means we assume $\partial|a\rangle/\partial R=\partial^{2}|a\rangle/\partial R^{2}=0$.
From this approximation the time-independent Schr\"odinger equation,
\[
\hat{H}|\Psi\rangle=E|\Psi\rangle,
\]
can be manipulated to yield the multichannel equations for the radial
wavefunctions:
\begin{equation}
\sum_{a}\left[\frac{-\hbar^{2}}{2\mu}\frac{d^{2}}{dR^{2}}\delta_{a'a}+V_{a'a}(R)-E\delta_{a'a}\right]G_{a}(R)=0,\label{eq:multichannel equations}
\end{equation}
where $V_{a'a}(R)=\langle a'|\hat{V}(R)|a\rangle$ and $\hat{V}=\hat{H}-\hat{T}$.

\subsection{Basis}

It is difficult to directly write down a basis that simultaneously
diagonalises the Zeeman and Hyperfine interactions which are both
present at large separation distances. Instead we perform the calculation
in a basis whose elements can be easily listed and find the eigenstates
by numerically diagonalising the long-distance Hamiltonian.

The easily enumerated basis we use is the hyperfine basis:
\[
|a\rangle\equiv\text{|\ensuremath{\alpha\beta\,lm_{l}\rangle\equiv|(S_{\alpha}i_{\alpha}f_{\alpha}m_{f_{\alpha}})(S_{\beta}i_{\beta}f_{\beta}m_{f_{\beta}})\,lm_{l}\rangle}}.
\]
Here $\alpha$ and $\beta$ list the magnetic spin numbers for atoms
$\alpha$ and $\beta$, with the electronic and nuclear spins of each
atom coupled as $\mathbf{S}_{j}+\mathbf{i}_{j}=\mathbf{f}_{j}$. $l=0,1,2,\hdots$
is the rotational angular momentum of the two-atom system.

In the case of homonuclear collisions these basis states must be symmetrised
to obey the fermionic and bosonic symmetry of $\prescript{3}{}{}$He$^*$-$\prescript{3}{}{}$He$^*$
and $\prescript{4}{}{}$He$^*$-$\prescript{4}{}{}$He$^*$ respectively
\citep{stoofSpinexchangeDipoleRelaxation1988}. We form symmetrised
states $|\{\alpha\beta\}\,lm_{l}\rangle$ as linear combinations \citep{cocksUltracoldHomonuclearHeteronuclear2019}:
\[
|\{\alpha\beta\}\,lm_{l}\rangle=\frac{|\alpha\beta\rangle+(-1)^{i_{\alpha}+i_{\beta}+l}|\beta\alpha\rangle}{\sqrt{2(1+\delta_{\alpha\beta})}}\otimes|lm_{l}\rangle.
\]
In the case of $\prescript{3}{}{}$He$^*$-$\prescript{4}{}{}$He$^*$
collisions we use the unsymmetrised $|\alpha\beta\,lm_{l}\rangle$
states.

\subsection{Matrix elements\label{subsec:Matrix-elements}}

When integrating the multichannel equations (\ref{eq:multichannel equations})
we must calculate the matrix elements $V_{aa'}(R)$.

The rotational interaction has the simplest matrix element:
\[
\langle a'|\hat{H}_{\mathrm{rot}}|a\rangle=\delta_{a,a'}\frac{l(l+1)\hbar^{2}}{2\mu R^{2}}.
\]

The matrix elements for $\hat{H}_{\mathrm{el}},$ $\hat{H}_{\mathrm{zee}}$,
and $\hat{H}_{\mathrm{sd}}$ are evaluated by changing into the basis
associated with the coupling scheme $\mathbf{S_{\alpha}+S_{\beta}=S}$,
using the expansion in terms of Clebsch-Gordan coefficients:
\begin{align*}
|\alpha\beta\rangle & =\sum_{\substack{m_{S}^{\alpha}\\
m_{i}^{\alpha}
}
}\sum_{\substack{m_{S}^{\beta}\\
m_{i}^{\beta}
}
}\sum_{\substack{S\\
m_{S}
}
}C_{m_{S}^{\alpha}m_{i}^{\alpha}m_{f}^{\alpha}}^{S_{\alpha}i_{\alpha}f}C_{m_{S}^{\beta}m_{i}^{\beta}m_{f}^{\beta}}^{S_{\beta}i_{\beta}f_{\beta}}C_{m_{S}^{\alpha}m_{S}^{\beta}m_{S}}^{S_{\alpha}S_{\beta}S}\\
 & \times|(S_{\alpha}S_{\beta}Sm_{S})(i_{\alpha}m_{i}^{\alpha}i_{\beta}m_{i}^{\beta})\rangle.\tag{{\theequation}}\stepcounter{equation}
\end{align*}

In this basis $\hat{H}_{\mathrm{el}}$ is diagonal: 
\[
\hat{H}_{\mathrm{el}}|Sm_{S}\rangle=^{2S+1}V_{\Sigma}(R)|Sm_{S}\rangle.
\]
 The quintet Born-Oppenheimer $^{5}\Sigma_{g}^{+}$ potential is taken
as the analytic form described by Przybytek and Jeziorski \citep{przybytekBoundsScatteringLength2005}.
The singlet $^{1}\Sigma_{g}^{+}$ and triplet $^{3}\Sigma_{u}^{+}$
potentials are interpolated from the tabulated values of M\"uller
et al. \citep{mullerExperimentalTheoreticalStudies1991} within the
$3\,a_{0}<R<14\,a_{0}$ range where those values are given, and fitted
to the $^{5}\Sigma_{g}^{+}$ potential at larger distances. Following
the method of \citep{cocksUltracoldHomonuclearHeteronuclear2019},
this fitting is done using an exponentially decaying exchange term:
$\prescript{1,3}{}{V_{\Sigma}}(R>14\,a_{0})=\prescript{5}{}{V_{\Sigma}}(R)-A_{1,3}\exp(-\beta_{1,3}R)$,
where $A_{1}=5.9784,\,\beta_{1}=0.7367,\,A_{3}=1.7980,$ and $\beta_{3}=0.6578$.

We write the spin-dipole as the product of two second-rank irreducible
tensors $\hat{H}_{\mathrm{sd}}=V_{p}(R)\mathbf{T}^{2}\cdot\mathbf{C}^{2}$,
where $V_{p}(R)=b/R^{3}$ and $b=-\sqrt{6}\xi$ \citep{beamsSpindipoleinducedLifetimeLeastbound2006}.
This gives the matrix elements as
\[
\langle a'|\hat{H}_{\mathrm{sd}}|a\rangle=V_{p}(R)D_{aa'}.
\]
 The coupling coefficient is
\begin{align*}
D_{aa'} & =\delta_{m_{S'}+m_{l'},m_{S}+m_{l}}(-1)^{m_{S}'-m_{S}}\\
 & \times C_{m_{S'}\,m_{S'}-m_{S}\,m_{S'}}^{S\,2\,S'}C_{m_{l}\,m_{l'}-m_{l}\,m_{l'}}^{l\,2\,l'}\\
 & \times\langle S_{\alpha}'S_{\beta}'S'||\mathbf{T}^{2}||S_{\alpha}S_{\beta}S\rangle\langle l'||\mathbf{C}^{2}||l\rangle.\tag{{\theequation}}\stepcounter{equation}
\end{align*}
The reduced matrix elements for the tensors $\mathbf{T}^{2}$ and
$\mathbf{C}^{2}$ are
\begin{align*}
\langle S_{\alpha}'S_{\beta}'S'||\mathbf{T}^{2}||S_{\alpha}S_{\beta}S\rangle & =\delta_{S_{\alpha},S_{\alpha}'}\delta_{S_{\beta},S_{\beta}'}\\
 & \times\sqrt{S_{\alpha}(S_{\alpha}+1)S_{\beta}(S_{\beta}+1)}\\
 & \times\sqrt{5(2S+1)(2S_{\alpha}+1)(2S_{\beta}+1)}\\
 & \times\begin{Bmatrix}S_{\alpha} & S_{\beta} & S\\
1 & 1 & 2\\
S_{\alpha} & S_{\beta} & S'
\end{Bmatrix},\tag{{\theequation}}\stepcounter{equation}
\end{align*}
where the last factor is a Wigner 9-j coefficient, and
\[
\langle l'||\mathbf{C}^{2}||l\rangle=\sqrt{\frac{2l+1}{2l'+1}}C_{0\,0\,0}^{l\,2\,l'}.
\]

The matrix elements for the Zeeman interaction are simply
\[
\langle S'\,m_{S'}|\hat{H}_{\mathrm{zee}}|S\,m_{S}\rangle=\delta_{S,S'}\delta_{m_{S},m_{S'}}\frac{g_{s}\mu_{B}}{\hbar}Bm_{S}.
\]
The relative difference between the electron spin g-factor and the
$2\prescript{3}{}{S}_{1}$ g-factor is of the order of $10^{-5}$
\citep{drakeMagneticMomentHelium1958} and we therefore neglect it.
We also neglect the contribution from nuclear spin because the nuclear
magnetic moment is on the order of $10^{3}$ smaller than the electronic
magnetic moment \citep{flowersMeasurementNuclearMagnetic1993}.

$\hat{H}_{\mathrm{hfs}}$ is diagonal in this basis, with the matrix
elements \citep{rosnerHyperfineStructure3S11970}:
\[
\langle a'|\hat{H}_{\mathrm{hfs}}|a\rangle=\delta_{a,a'}(E_{i_{\alpha},f_{\alpha}}^{\mathrm{hfs}}+E_{i_{\beta},f_{\beta}}^{\mathrm{hfs}})
\]
where
\begin{align*}
E_{i,f}^{\mathrm{hfs}} & =\begin{cases}
1.519830\times10^{-7}\,E_{\mathrm{h}} & i=\frac{1}{2},f=\frac{3}{2}\\
0\,E_{\mathrm{h}} & \mathrm{otherwise}
\end{cases}.\tag{{\theequation}}\stepcounter{equation}
\end{align*}

To optimise the numerical integration routine we recognise that the
non-diagonal interactions---$\hat{H}_{\mathrm{el}}$, $\hat{H}_{\mathrm{sd}}$
and $\hat{H}_{\mathrm{zee}}$---can be written separably as products
of $R$-independent matrices and a scalar functions of $R$. For example
in the case of the electronic term,
\[
\langle a|\hat{H}_{\mathrm{el}}(R)|a'\rangle=\sum_{S=0}^{2}C_{aa'}^{2S+1}V^{2S+1}(R),
\]
where each $C_{aa'}^{2S+1}$ is a coupling coefficient. This separation
of R-independent coupling terms can be written more generally as:
\[
V_{aa'}(R)=\sum_{k}V_{aa'}^{k}\otimes f_{k}(R),
\]
where $V_{aa'}^{k}$ would be a coupling coefficient and $f_{k}$
a radial function. There are three different radial independent terms
for $\hat{H}_{\mathrm{el}}$ (corresponding to the three values of
$S=0,1,2$) and one each for $\hat{H}_{\mathrm{sd}}$ and $\hat{H}_{\mathrm{zee}}$.
We pre-calculate the radial-independent coupling coefficients for
the non-diagonal interactions so that only the radial factors are
repeatedly calculated.

\subsection{Scattering Matrices and Cross sections}

The scattering problem is constrained by inner and outer boundary
conditions as $R\rightarrow0$ and $R\rightarrow\infty$ respectively.
All radial wavefunctions must vanish at the inner boundary where the
potential diverges; additionally, all closed channel wavefunctions
must vanish at the outer boundary. We define linearly independent
initial conditions representing both sets of boundary conditions,
and integrate those conditions to a single middle point. We then use
a QR decomposition to find linear combinations of the boundary conditions
that produce matching solutions at that point. These matched solutions
satisfy both boundary conditions.

The matched solutions, evaluated at the outer boundary where the electronic
and spin-dipole interactions can be safely neglected, are then fitted
to spherical bessel functions, which represent the oscillatory asymptotic
form of the scattering wave. Specifically, the solutions $\mathbf{F}(R)$
are matched to the form \citep{miesScatteringTheoryDiatomic1980}:
\[
\mathbf{F}(R)\xrightarrow[R\rightarrow\infty]{}\mathbf{J}(R)\mathbf{A}+\mathbf{N}(R)\mathbf{A},
\]
where $\mathbf{J}(R)$ and $\mathbf{N}(R)$ are diagonal matrices
with the entries
\begin{align*}
J_{aa}(R) & =\sqrt{k_{a}R}j_{l_{a}}(k_{a}R),\\
N_{aa}(R) & =\sqrt{k_{a}R}n_{l_{a}}(k_{a}R).\tag{{\theequation}}\stepcounter{equation}
\end{align*}
The matrices have one entry for each open channel $a$.

The fitted matrices $\mathbf{A}$ and $\mathbf{B}$ are used to define
the reactance matrix $\mathbf{K=\mathbf{BA}}^{-1},$ which then defines
the scattering matrix:
\[
\mathbf{S}=(\mathbf{I}+i\mathbf{K})(\mathbf{I}-i\mathbf{K})^{-1}.
\]

Cross sections for scattering from one channel into another (including
elastic scattering where the incoming and outgoing channels are identical)
are defined from the scattering matrix: 
\[
\sigma(\gamma\rightarrow\gamma')=\frac{\pi}{k_{\gamma}^{2}}\sum_{\substack{l\,m_{l}\\
l'\,m_{l}'
}
}|T_{\gamma'l'm_{l}',\,\gamma lm_{l}}|^{2},
\]
where the transition matrix $\mathbf{T=\mathbf{I}-\mathbf{S}}$.

Similarly, the ionisation cross sections are found from the non-unitarity
of the scattering matrix:

\[
\sigma(\gamma\rightarrow\mathrm{PI})=\sum_{l\,m_{l}}\left[1-\sum_{\gamma'l'm_{l}'}|S_{\gamma'l'm_{l}',\,\gamma lm_{l}}|^{2}\right].
\]

\subsection{Resonance search}

Elastic and ionisation cross sections are calculated at constant collisional
energy over a grid of magnetic fields ranging from $0\,\mathrm{G}$
to $10^{4}\,\mathrm{G}$ ($0\,\mathrm{T}$ to $1\,\mathrm{T}$). A
maximum grid spacing of $400\,\mathrm{G}$ is used to observe the
broad behaviour and identify resonances, with smaller grid spacing
used around features of interest and where resonances were predicted
in \citep{goosenFeshbachResonances3He2010}.

For each isotopic mixture we consider the two most experimentally
relevant scattering channels, which are the two channels with lowest
energy (under the hyperfine and Zeeman interactions). In bosonic He$^*$
the lowest energy channel corresponds to spin polarised atoms in the
$m_{f}=+1$ state---this channel is the most experimentally accessible
because the polarisation minimises Penning ionisation \citep{shlyapnikovDecayKineticsBose1994}.
We do not expect resonances in the lowest energy fermionic channel
as that channel involves identical fermions. However, we consider
resonances in the second lowest energy fermionic channel, which can
be experimentally accessed for example by using a radiofrequency sweep
to transfer atoms from the lowest energy $|f,m_{f}\rangle=|3/2,3/2\rangle$
state to the $|3/2,1/2\rangle$ state before collision.

From the elastic scattering cross sections we calculate resonance
locations and widths, by fitting Fano profiles to the data according
to the equation:
\[
\sigma_{\mathrm{el}}^{\mathrm{Fano}}(B)=\sigma_{0}\frac{(q\frac{\Delta}{2}+B-B_{0})^{2}}{\left(\frac{\Delta}{2}\right)^{2}+(B-B_{0})^{2}}.
\]
Here $\sigma_{0}$ is the non-resonant background cross section, $B_{0}$
is the resonance location, $q$ is a shape parameter, and $\Delta$
is the resonance width.

\section{Results and Discussion}

\begin{table}[bt]
\caption{Identified resonance positions $B_{0}$ and widths $\Delta$. Scattering
channels are labelled in order of their energy: A and B respectively
correspond to the lowest and second-lowest energy collision channels.
In $\prescript{4}{}{}$He$^*$ the A channel corresponds to spin polarised
atoms in the $m_{f}=+1$ state. When $\prescript{3}{}{}$He$^*$ atoms
are involved the channels are linear combinations of different hyperfine
states. Uncertainty in $B_{0}$ was calculated by determining the difference
between the location of the $9.75(2)\times10^{3}\,\mathrm{G}$ resonance
and twice the location of the $4881(9)\,\mathrm{G}$ resonance. Uncertainty
in $\Delta$ was taken as the fitting uncertainty given when fitting
Fano profiles to the data.}
\begin{center}\begin{threeparttable}\label{table of resonances}

\begin{tabular}{cccc}
\toprule 
Mixture & Scattering channel & $B_{0}$ (G) & $\Delta$ (G)\tnote{b}\tabularnewline
\midrule
\midrule 
44 & A & 345.0(7)\tnote{a} & -\tabularnewline
\midrule 
 & A & 4881(9) & 0.05(1)\tabularnewline
\midrule 
 & A & $9.75(2)\times10^{3}$ & -\tabularnewline
\midrule 
34 & A & $130.0(2)$\tnote{a} & -\tabularnewline
\midrule 
 & A & 4749(9) & -\tabularnewline
\midrule 
 & A & $9.55(2)\times10^{3}$ & -\tabularnewline
\midrule 
33 & B & $1.760(3)\times10^{4}$ & 0.05(2)\tabularnewline
\bottomrule
\end{tabular}

\begin{tablenotes}
\item [a] Resonance observable only when ionisation is switched off.
\item [b] Widths are not given where the fit to the Fano profile produced errors greater than the widths themselves.
\end{tablenotes}
\end{threeparttable}\end{center}
\end{table}

Seven resonances identified in this study are described in Table~\ref{table of resonances}.
Three each are identified in the lowest energy scattering channels
of heteronuclear and bosonic collisions. Another resonance is identified
in the second-lowest energy scattering channel of fermionic collisions.
In this discussion we compare with the results of \citep{goosenFeshbachResonances3He2010},
which is the only existing theoretical study of He$^*$ Feshbach resonances.

\begin{figure}[tb]
\includegraphics[scale=0.27]{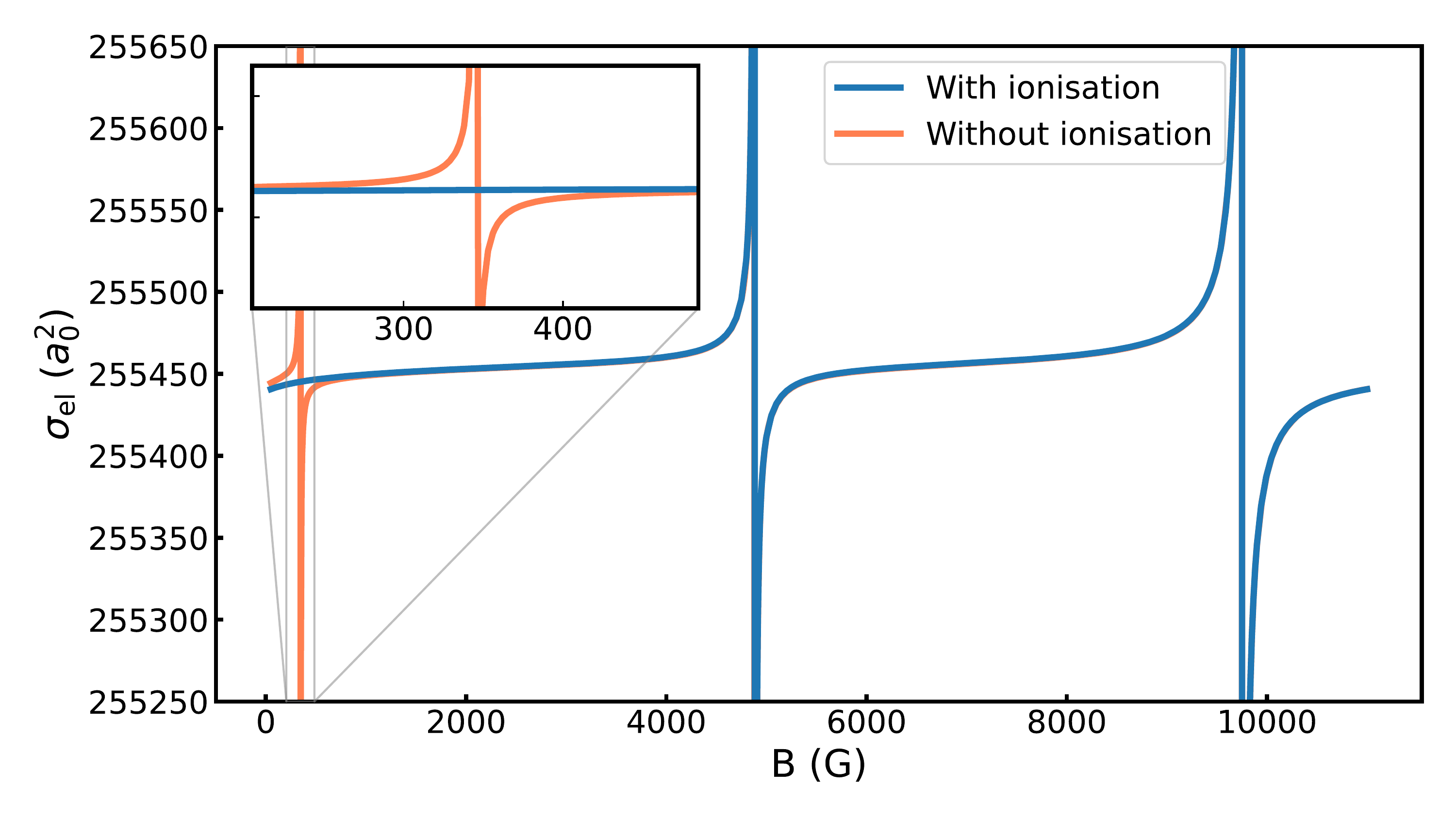}

\caption{Elastic cross section of bosonic $\prescript{4}{}{}$He$^*$-$\prescript{4}{}{}$He$^*$
collisions in the lowest energy scattering channel. Blue shows cross
sections with Penning ionisation accounted for; orange shows cross
sections with Penning ionisation neglected. The orange curve lies
under the blue curve except for the first resonance.}

\label{44-aa-res}
\end{figure}

Two of the three resonances identified in homonuclear $\prescript{4}{}{}$He$^*$
collisions (rightmost two resonances in Figure~\ref{44-aa-res})
roughly coincide with the previous theoretical study. The resonance
at $345.0(7)\,\mathrm{G}$ is located within the range of previously forecast locations,
which was $99\,\mathrm{G}$ to $460\,\mathrm{G}$ \citep{goosenFeshbachResonances3He2010}.
This resonance has been investigated experimentally but not found
\citep{borbelyMagneticfielddependentTrapLoss2012}. However, upon switching off ionisation in our model (blue line in Figure~\ref{44-aa-res}), a matching resonance appears. When
the ionisation effects are included (orange line in in Figure~\ref{44-aa-res}), our results indicate that it is broadened massively by ionisation processes, which were neglected in the prior
study. These results offer a complete explanation of the discrepancy
between theory and experiment.

We believe the resonance seen at $4881(9)\,\mathrm{G}$ in Figure~\ref{44-aa-res}
is the same as a previously predicted resonance at $5460\,\mathrm{G}$
\citep{goosenFeshbachResonances3He2010}. Single channel calculations
of bound state energies via counting of wavefunction nodes \citep{messiahQuantumMechanics1999},
performed using this model, matched the resonance seen here to the
same associated bound state as in the previous prediction. The third
resonance at $9.75(2)\times10^{3}\,\mathrm{G}$ was not previously
predicted, however its appearance is unsurprising. The $4881(9)\,\mathrm{G}$
resonance is caused by coupling between the $|S=2,m_{S}=+2\rangle$
scattering channel and a bound state in the $|S=2,m_{S}=0\rangle$
closed channel; however, that scattering channel may also couple to
a bound state in the $|S=2,m_{S}=+1\rangle$ channel. Because the
Zeeman potential is directly proportional to the spin projection,
when the spin projection between the channels is halved, the resonance
condition for the same bound state:
\[
\frac{-g_{S}\mu_{B}}{\hbar}B_{0}(m_{S_{\mathrm{scat}}}-m_{S_{\mathrm{bound}}})=E_{\mathrm{bound}}-\frac{\hbar^{2}k^{2}}{2\mu},
\]
 occurs at twice the magnetic field. Indeed, $9.75(2)\times10^{3}\,\mathrm{G}$
is within $0.1\%$ of $2\times4881\,\mathrm{G}$. We in fact used
this difference (between our model data and the precise factor of 2) to
estimate the relative uncertainty in $B_{0}$, giving an uncertainty
estimate which we propagated to all other resonances in our model.
Note that this estimate does not reflect uncertainty from our use of the
Born-Oppenheimer potentials, which themselves are uncertain---particularly
the singlet and triplet potentials of \citep{mullerExperimentalTheoreticalStudies1991}.

\begin{figure}[t]
\includegraphics[scale=0.27]{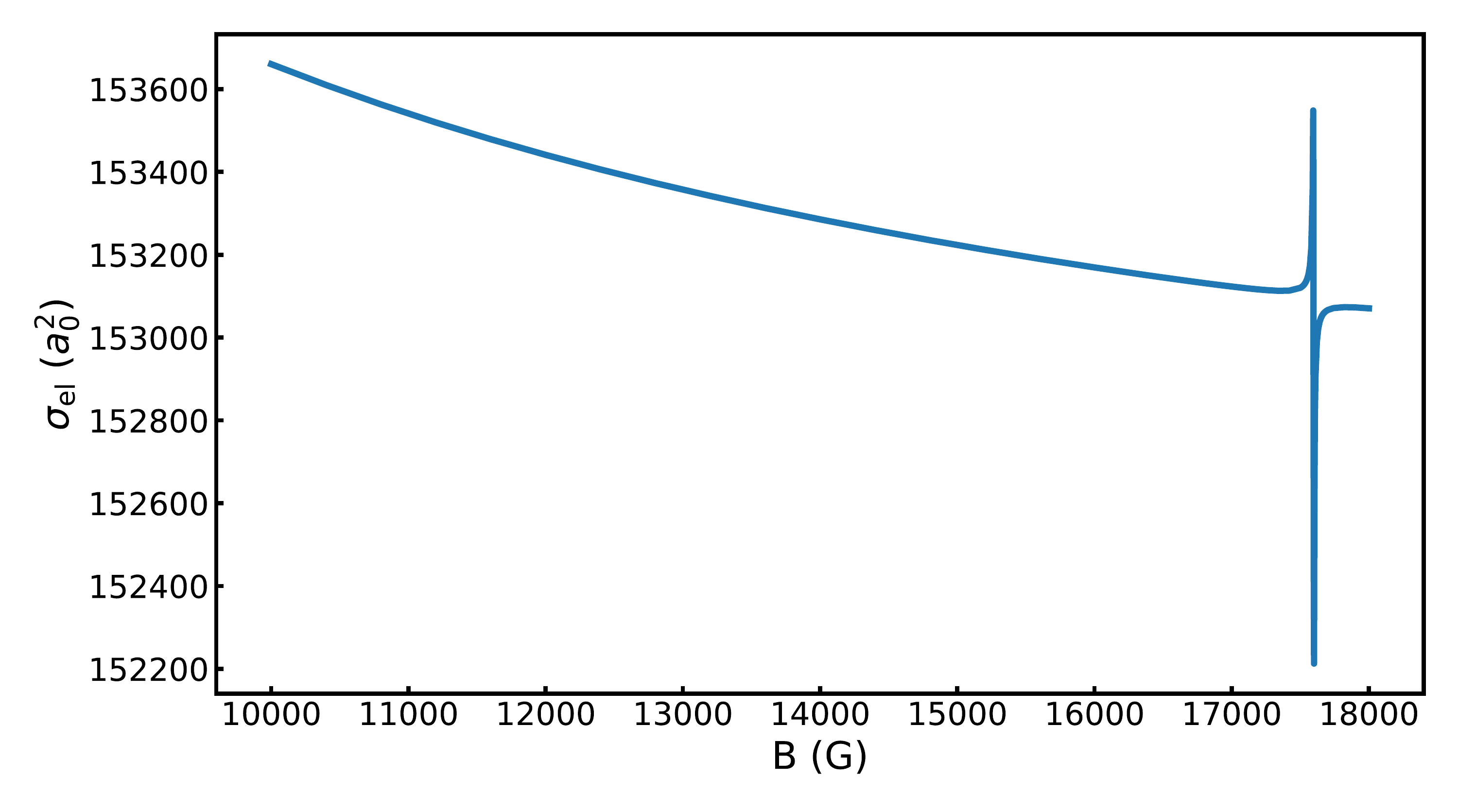}

\caption{Elastic cross section of fermionic $\prescript{3}{}{}$He$^*$-$\prescript{3}{}{}$He$^*$
collisions in the second-lowest energy scattering channel. A single
Feshbach resonance is identified near $17600\,\mathrm{G}$. This scattering
channel is a linear combination of states where spin polarised atoms
collide with unpolarised atoms of $m_{f}=-1/2$.}

\label{33-ab}
\end{figure}

The fermionic Feshbach resonance seen in Figure~\ref{33-ab}
was also predicted previously \citep{goosenFeshbachResonances3He2010},
with our calculated value at the upper bound of the previously predicted
range of values. Those previous bounds were determined by variation
in the $S=1$ potential.

\begin{figure}[hbt]
\includegraphics[scale=0.27]{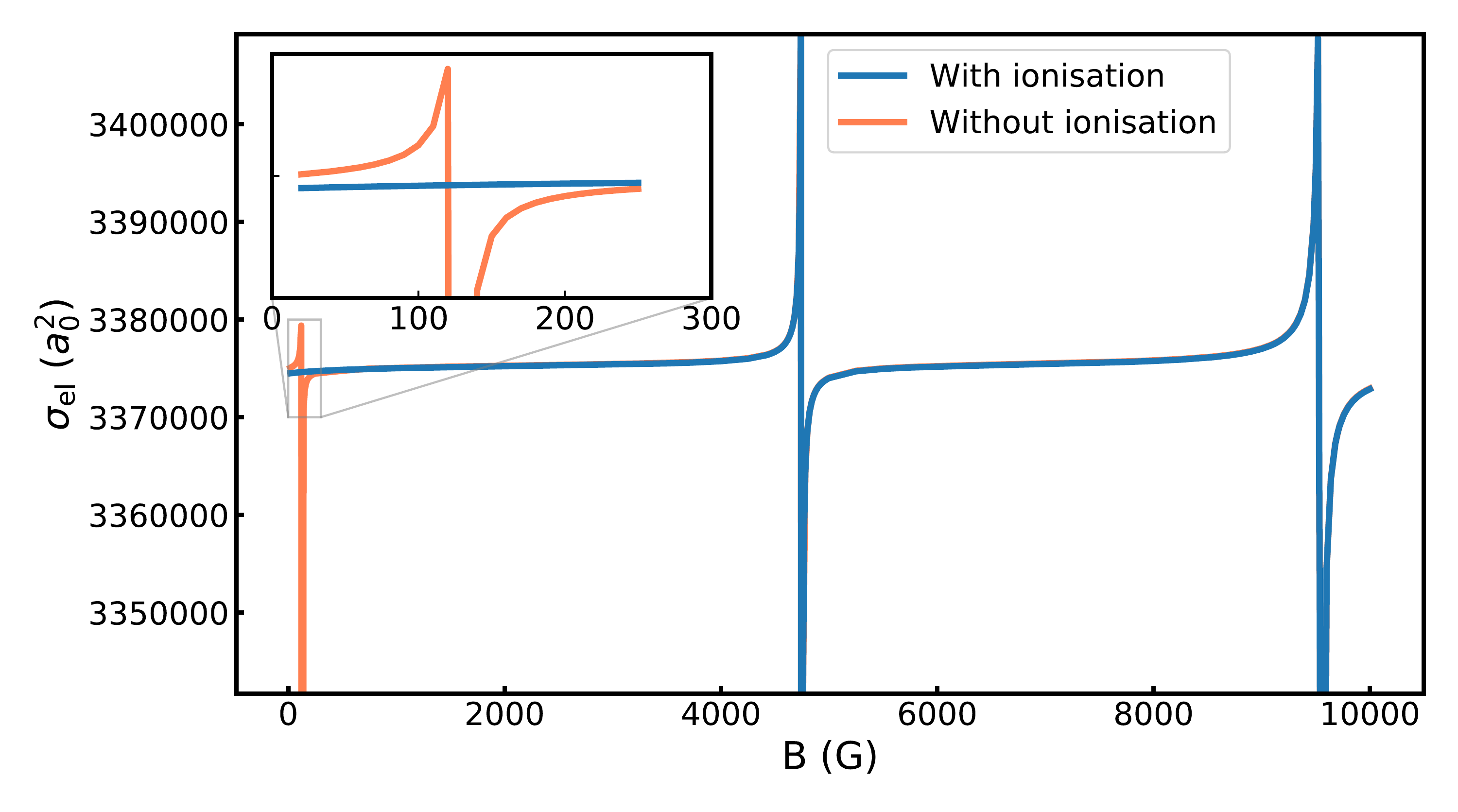}

\caption{Elastic cross section of heteronuclear $\prescript{3}{}{}$He$^*$-$\prescript{4}{}{}$He$^*$
collisions in the lowest energy scattering channel. Blue shows cross
sections with Penning ionisation accounted for; orange shows cross
sections with Penning ionisation neglected. Inset: one of the resonances
is only visible when ionisation is discounted from the model. The
orange curve lies under the blue curve except for near the location of the first resonance.}

\label{34-aa-res}
\end{figure}

The heteronuclear A channel resonance structure (Figure~\ref{34-aa-res})
is qualitatively very similar to the bosonic A channel structure (as shown in Figure~\ref{44-aa-res}).
This is likely due to the similarity between the homonuclear bosonic
system and the reduced set of heteronuclear channel states that are
directly coupled by the Zeeman interaction. Both of the larger two
heteronuclear resonances were previously predicted to exist \citep{goosenFeshbachResonances3He2010}.
Interestingly, the resonance centred at $9.55(2)\times10^{3}\,\mathrm{G}$
was predicted to be narrower than the $4749(6)\,\mathrm{G}$ resonance,
but in these results resonance widths increase as $B_{0}$ increases.
As with the small bosonic resonance, the ionisation process broadens
the $130\,\mathrm{G}$ resonance to the extent that it cannot be seen.

\begin{figure}[t!]
\includegraphics[scale=0.27]{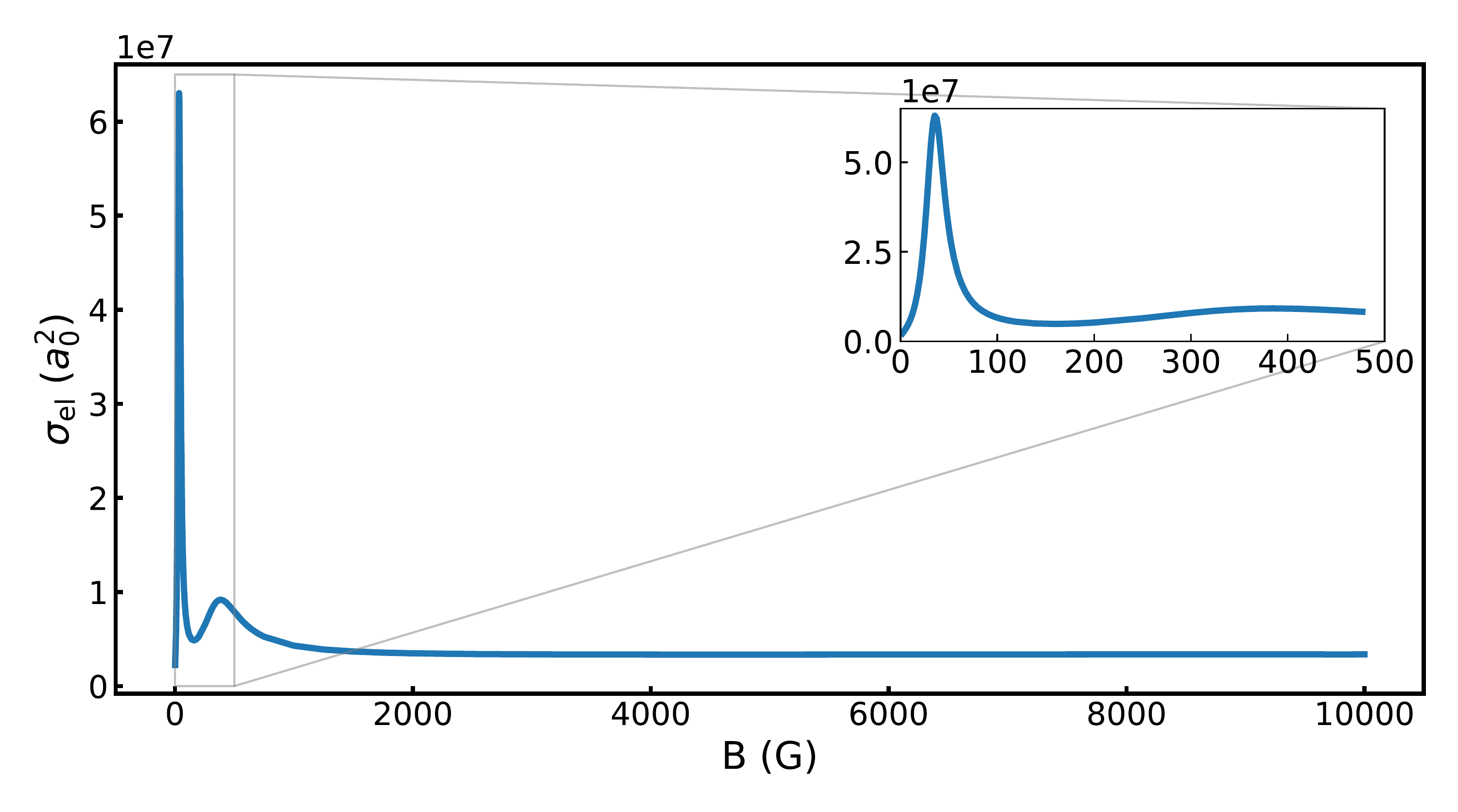}
\caption{Elastic cross section of heteronuclear $\prescript{3}{}{}$He$^*$-$\prescript{4}{}{}$He$^*$
collisions in the second-lowest energy scattering channel. Inset:
while no Fano profiles indicative of Feshbach resonances are identified,
the cross section varies by over an order of magnitude across an experimentally
accessible range of magnetic fields weaker than $100\,\mathrm{G}$.}

\label{34-ab-var}
\end{figure}

No Fano profiles were observed in the heteronuclear B channel (Figure~\ref{34-ab-var}),
not even a very wide resonance previously predicted at $1214\,\mathrm{G}$.
Three other resonances in this channel were previously predicted,
however without resonance widths \citep{goosenFeshbachResonances3He2010}.
We thoroughly searched for a resonance predicted at $3618(6)\,\mathrm{G}$,
scanning from $3598\,\mathrm{G}$ to $3638\,\mathrm{G}$ with a grid
spacing of $1\,\mathrm{G}$, however no field dependence was identified
in the cross section.

Interestingly, Figure~\ref{34-ab-var} shows that in
this scattering channel there are variations in the elastic cross
section that are inconsistent with a Fano profile. Similar curves
appear in the ionisation cross sections (not shown). The sharp spike
in cross section between $0\,\mathrm{G}$ and $100\,\mathrm{G}$---reminiscent
of a shape resonance \citep{joachainQuantumCollisionTheory1975}---is
particularly interesting, with the elastic cross section there varying
by over an order of magnitude. This magnetic field range is experimentally
accessible, so while not being a true Feshbach resonance (i.e. with
a diverging scattering length), this behaviour could allow nontrivial
manipulation of the scattering length of heteronuclear He$^*$ collisions,
opening up several experimental possibilities. Experimental verification
of this variation is an interesting and open question.

Unfortunately, our results do not identify a Feshbach resonance that
appears experimentally useful. In the bosonic A channel ionisation
processes broaden the $345\,\mathrm{G}$ resonance to the point of
insignificance. The resonance at $4881(9)\,\mathrm{G}$ would likely
pose difficulties in an experiment, as its location and width would
demand a very powerful but stable electromagnet; these difficulties
are magnified for the $9.55(2)\times10^{3}$ resonance. This state
of affairs also describes the heteronuclear A channel where the resonance
structure is similar. The singular resonance in the fermionic B channel
occurs at an even larger magnetic field and is therefore even less useful.

In contrast to the identified (unbroadened) Feshbach resonances, the
non-Feshbach variation in the 3-4 B channel cross section occurs over
a very accessible range of magnetic fields. Because this variation
is not a Feshbach resonance the scattering length may not diverge.
However, with the elastic cross section varying by a factor of $\sim30$,
by $\sigma\approx4\pi a^{2}$ that would imply the scattering length
varies by a factor of $\sim5$, which is significant. Ionisation would
likely be a hindrance practically however, as our model calculates the ionisation cross
section for heteronuclear B channel collisions to be $\sim10^{7}$ times
larger than for spin polarised bosonic helium.

\section{Conclusion}

We developed a new close coupled model of ultracold He$^*$ atom collisions
in the presence of a magnetic field and used it to search for Feshbach
resonances. Several resonances were identified across bosonic, fermionic,
and heteronuclear scattering channels, however none are likely to
be experimentally accessible. We identified that a predicted Feshbach
resonance that was not seen in experiment is actually broadened by
ionisation processes. We also identified magnetic field dependence
in the cross section of the second-lowest energy heteronuclear scattering
channel, which although not a Feshbach resonance, occurs at low magnetic
fields that may be experimentally accessible.

\begin{acknowledgments}
The authors would like to thank James Sullivan and Andrew Truscott
for helpful discussions, and Andrew Truscott for careful reading of the manuscript. This work was supported through Australian
Research Council (ARC) Discovery Project grant DP190103021. DGC was
supported by ARC Discovery Early Career Researcher Award DE170101024.
\end{acknowledgments}

\bibliographystyle{unsrtnat}
\bibliography{bib}

\end{document}